\documentclass{nature2}
\usepackage{graphicx}
\usepackage[usenames]{color}
\usepackage{xcolor}
\usepackage[export]{adjustbox}

\usepackage{amsmath}
\usepackage{amssymb}

\title{Truchet-tile structure of a topologically aperiodic metal--organic framework}

\author{Emily G. Meekel,$^1$ Ella M. Schmidt,$^{1,2}$ Lisa J. Cameron,$^3$ A. David Dharma,$^3$ Hunter J. Windsor,$^3$ Samuel G. Duyker,$^{3,4}$ Arianna Minelli,$^1$ Tom Pope,$^5$ Giovanni Orazio Lepore,$^6$ Ben Slater,$^5$ Cameron J. Kepert$^{3,\ast}$ and Andrew L. Goodwin$^{1,\ast}$}

\begin{document}

\maketitle

\begin{affiliations}
 \item Inorganic Chemistry Laboratory, University of Oxford, Oxford OX1 3QR, UK
 \item Fachbereich Geowissenschaften, Universität Bremen, D-28359 Bremen, Germany
 \item School of Chemistry, University of Sydney, NSW 2006, Australia
 \item Sydney Analytical, Core Research Facilities, University of Sydney, NSW 2006, Australia
 \item Department of Chemistry, University College London, London WC1H 0AJ, UK
 \item Earth Sciences Department, University of Florence, Florence 50121, Italy
\end{affiliations}

\begin{abstract}

Periodic tilings can store information if individual tiles are decorated to lower their symmetry. Truchet tilings---the broad family of space-filling arrangements of such tiles---offer an efficient mechanism of visual data storage related to that used in barcodes and QR codes. Here, we show that the crystalline metal--organic framework [OZn$_{\mathbf 4}$][1,3-benzenedicarboxylate]$_{\mathbf 3}$ (TRUMOF-1) is an atomic-scale realisation of a complex three-dimensional Truchet tiling. Its crystal structure consists of a periodically-arranged assembly of identical zinc-containing clusters connected uniformly in a well-defined but disordered fashion to give a topologically aperiodic microporous network. We suggest that this unusual structure emerges as a consequence of geometric frustration in the chemical building units from which it is assembled.

\end{abstract}

In 1704, S{\'e}bastian Truchet described a variety of visually appealing patterns generated by a single square tile painted differently either side of its diagonal.\cite{Truchet_1704} Truchet always arranged his tiles on the same square lattice, but obtained different patterns by varying the orientations of each tile. Truchet tilings have since been generalised to encompass any periodic covering of space (partial or complete) with one or more tiles that have been decorated to reduce their symmetry.\cite{Smith_1987} Breaking colour-inversion symmetry, for example, gives barcodes and QR codes as simple Truchet tilings in one and two dimensions, respectively: each strip or pixel encodes a binary state that is exploited in their application as machine-readable information stores. More complex Truchet tilings store information in aperiodic patterns of loops and networks in two or three dimensions [Figure 1].\cite{Browne_2008}

\begin{figure}
\begin{center} 
\includegraphics{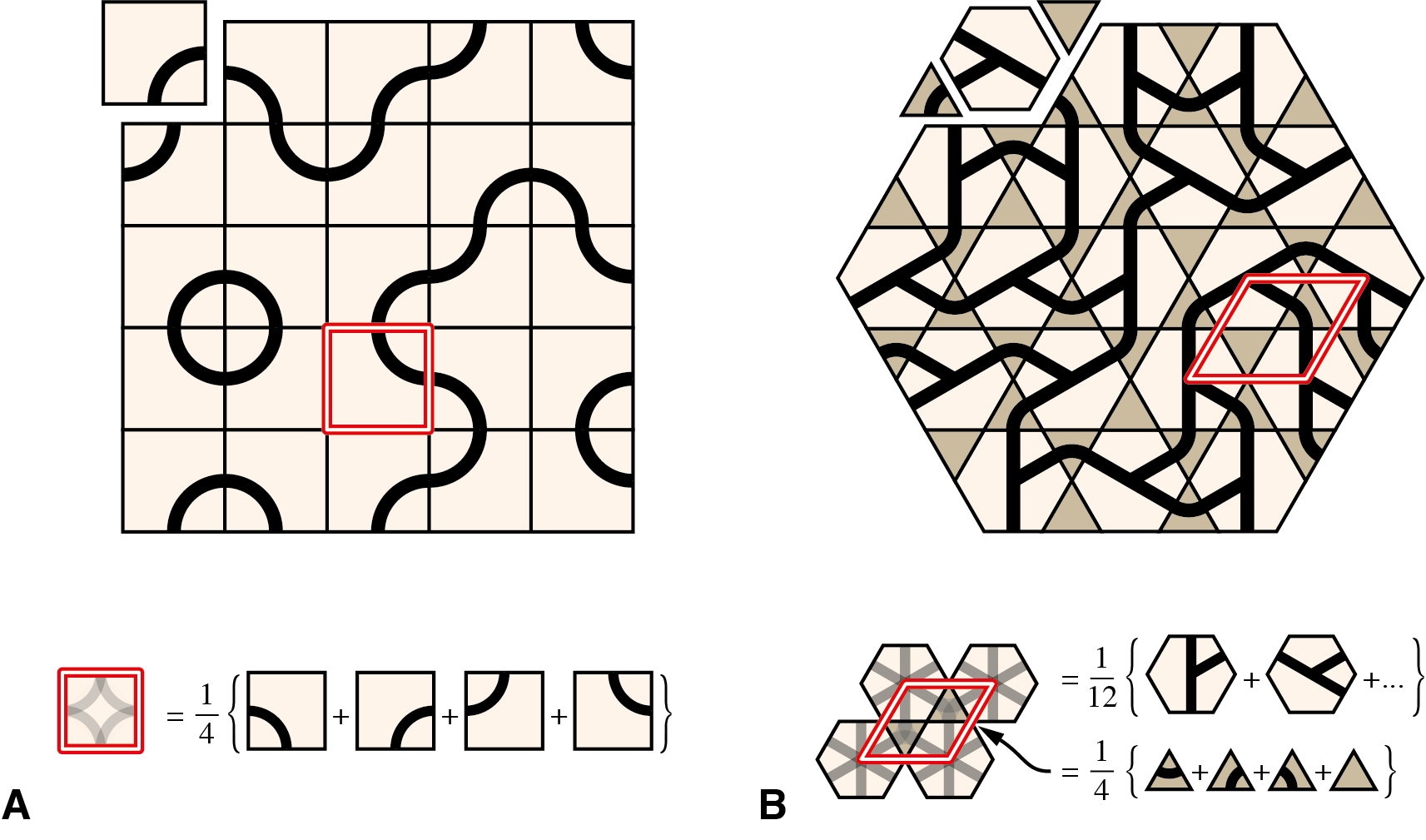}
\end{center}
\caption{\label{fig1}\footnotesize{\bf  Representative Truchet tilings. A} Decorating square tiles with a single arc can be used to generate Truchet tilings with curved paths. The positions of the square tiles are translationally periodic, but the continuous pattern produced by the arc decorations is aperiodic. Projecting any such tiling onto the underlying unit cell (shown in red) gives an average structure with partial occupancies. For large systems, one expects equal occupancies of each of the four possible arc orientations. {\bf B} More complex Truchet tilings can be generated using a combination of different decorated and undecorated tiles. In this particular example, based on the trihexagonal tiling of the plane, the connected path of line decorations forms a single percolating network that spans the entire system.}
\end{figure}

Because the solid-state structures of many crystalline materials can be related to periodic tilings of three-dimensional space,\cite{Blatov_2007} we were intrigued by the possibility of realising an atomic-scale Truchet tiling in one such system. Metal--organic frameworks (MOFs) are a particularly attractive family of materials in this context, because there is good collective understanding of how to control their structures through judicious selection of the chemical building blocks from which they are assembled.\cite{DelgadoFriedrichs_2007} In the canonical system MOF-5 [OZn$_4$][1,4-bdc]$_3$ (bdc = benzenedicarboxylate), for example, the combination of octahedrally-coordinating zinc-containing clusters and linear bridging linkers generates a structure with the simple cubic topology.\cite{Li_1999} This structure can be mapped to the tiling of space by cubes.\cite{OKeeffe_2014}

Inspired by the emerging concept of introducing complexity in MOFs through lowering the symmetry of their building blocks,\cite{Guillerm_2019,Meekel_2021} we synthesised a new derivative of MOF-5 in which the linear 1,4-bdc linker was replaced by its bent 1,3-bdc analogue. The crystalline material we obtained, which we call TRUMOF-1, gave an X-ray diffraction pattern that included both conventional Bragg diffraction reflections and very weak diffuse scattering. Because the underlying chemistries of TRUMOF-1 and MOF-5 are so similar, we anticipated that the composition of TRUMOF-1 should be [OZn$_4$][1,3-bdc]$_3$. Structure solution from the Bragg component (sensitive only to the configurational average) gives the result shown in Figure 2A: [OZn$_4$] clusters are positioned at the corners and face-centres of the $F\bar43m$ cubic unit cell, and are connected by what appears to be trigonally-symmetric linker molecules that are only partly occupied. Distinguishable in the Fourier maps at the linker sites are benzene C atoms and carboxylate C/O atoms, the occupancies of which refine to values of 0.70(4) and 0.46(3), respectively; these occupancies imply a crystal composition of [OZn$_4$][1,3-bdc]$_{2.8(2)}$ and are close to the values (0.75 and 0.5) expected for the idealised composition [OZn$_4$][1,3-bdc]$_3$. We note that the configurational averages of Truchet tilings necessarily contain features with partial occupancies [Figure 1]. TRUMOF-1 can be made in bulk: powder samples give X-ray diffraction patterns entirely consistent with our single-crystal measurements. The framework pores of as-synthesised samples contain solvent, which can be exchanged with other solvents and/or removed by heating under vacuum. The empty framework adsorbs a range of gases (including CO$_2$) and has a BET surface area of 765 m$^2$ g$^{-1}$. Its thermal stability is very similar to that of MOF-5, and its adsorption enthalpies are about twice as large. In all these various respects (covered in greater detail in the SI), TRUMOF-1 behaves as a conventional MOF.\cite{Eddaoudi_2000}

\begin{figure}
\begin{center} 
\includegraphics{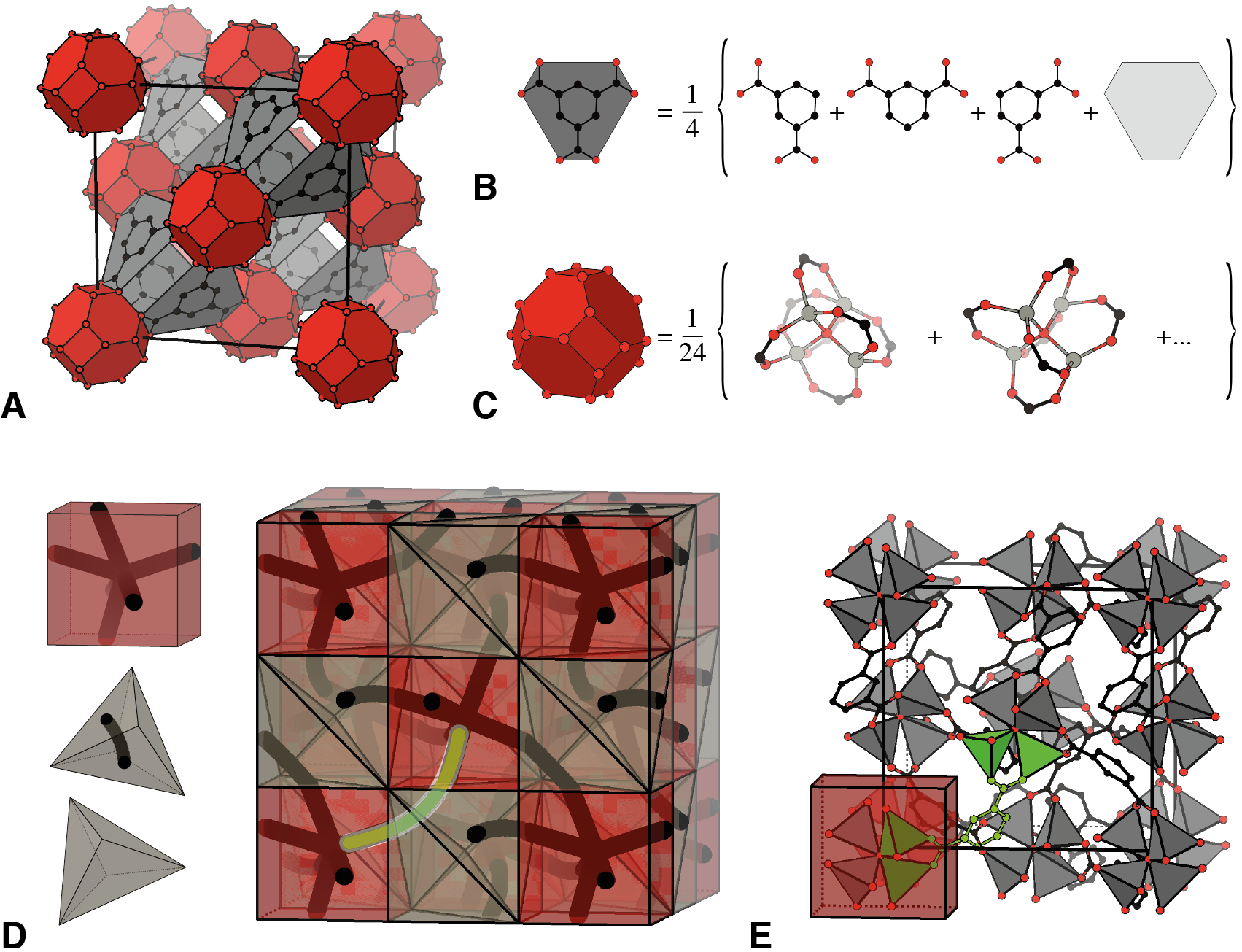}
\end{center}
\caption{\label{fig2}\footnotesize{\bf Crystal structure of TRUMOF-1. A} Representation of the average structure of TRUMOF-1 with OZn$_4$-centred polyhedra shown as red cuboctahedra and trigonally-symmetric ligand sites shown as grey triangles. {\bf B} Interpretation of the crystallographic model at the ligand site in terms of a configurational average over three 1,3-bdc orientations and 25\% linker vacancies. C atoms shown in black, O atoms in red, and H atoms omitted for clarity. {\bf C} Related interpretation of the inorganic cluster as an average over 24 octahedral cluster orientations. Colours are as in panel B, with Zn atoms shown in grey. {\bf D} The Truchet tiles used in our mapping correspond to (top--bottom) the pseudo-octahedral inorganic cluster orientations, the 1,3-bdc linker orientations, and the positions of ligand vacancies. Connecting these tiles to fill three-dimensional space forms networks that---like the patterns in Figure 1---are aperiodic. Note that the trigonal bipyramids pack such that any four combine to form a cube of the same dimensions as the (red) cluster cube. {\bf E} One possible $1\times1\times1$ approximant of the TRUMOF-1 structure with atom coordinates and cell geometry optimised using DFT. Note that the cluster positions are as in panel A, but that not all linker sites are occupied. The relationship between cluster and cubic Truchet tile is emphasised for one cluster; likewise the connectivity between two clusters in the foreground and the corresponding link in the related Truchet tiling (panel D) are highlighted in green.}
\end{figure}

How should we interpret the average crystal structure of TRUMOF-1? Considering first the partially-occupied ligand site, $^{1}$H/$^{13}$C NMR measurements of acid-digested TRUMOF-1 samples clearly showed the only bridging ligand present to be 1,3-bdc. So we interpret the crystallographic model to represent an average over three equally-populated orientations of the 1,3-bdc anion [Figure 2B]. The emergence of an apparently threefold-symmetric ligand is a consequence of the crystallographic site having higher point symmetry than the linker itself. To give the chemically sensible stoichiometry [OZn$_4$][1,3-bdc]$_3$, the site must also be vacant with 25\% probability; note that this value is consistent with the observed site occupancies. By contrast, the OZn$_4$ cluster appears ordered and is fully occupied. The cluster is surrounded by 12 carboxylate anion sites, each occupied with 50\% probability. Zn K-edge EXAFS measurements showed the Zn coordination environment to be essentially identical to that in the octahedrally-coordinated OZn$_4$ clusters of MOF-5,\cite{Hafizovic_2007} and so we interpret this aspect of the crystallographic model in terms of a configurational average over 24 symmetry-related octahedral-coordination decorations of the cluster [Figure 2C]. Again, each cluster decoration has lower point symmetry than the crystallographic site on which the cluster sits (the carboxylate positions break ideal octahedral symmetry; see SI for further discussion). One expects on purely geometric grounds the various partial site occupancies to be strongly correlated: the orientation of a given cluster decoration places strong constraints on which neighbouring 1,3-bdc linkers are present and in which possible orientations they lie.\cite{Simonov_2020,Ehrling_2021}

These ingredients of local symmetry-lowering and orientational matching rules allow a mapping to a three-dimensional Truchet tiling. We will come to test the validity of this interpretation against computation and experiment alike. To obtain the underlying tiling, we use a Voronoi decomposition of the crystal structure based on the average cluster and ligand positions. The two tile shapes so produced---a cube and a trigonal bipyramid---fill three-dimensional space when placed on the TRUMOF-1 cluster and ligand sites, respectively. The Truchet decorations are given by the symmetry-lowering implied by 1,3-bdc occupancies in the TRUMOF-1 crystal structure [Figure 2D]. So, for example, each cube is decorated by a pseudo-octahedral arrangement of rods that capture the arrangement of carboxylate substituents around each OZn$_4$ cluster. Likewise, three quarters of the trigonal bipyramids are decorated by an arc that reflects the bent linking geometry of a 1,3-bdc molecule. The remaining quarter of the trigonal bipyramidal tiles is left undecorated. Four triangular bipyramids pack together to form a cube. Any combination of orientations of these different tiles---alternating on their respective sites---that produces a connected network then relates to a chemically-sensible model of 1,3-bdc-linked OZn$_4$ clusters in which each cluster is octahedrally coordinated by six 1,3-bdc linkers and each 1,3-bdc linker connects exactly two clusters [Figure 2E]. There are infinitely many such networks.

To test whether this Truchet-tile description of TRUMOF-1 is physically reasonable, we carried out a series of \emph{ab initio} density functional theory (DFT) calculations. Our approach was to calculate the geometry and cell optimised 0\,K energies of periodic realisations (`approximants')\cite{Thygesen_2017} of the system with increasingly large unit cells so as to understand the underlying energetic landscape. There are only ten symmetry-distinct $1\times1\times1$ realisations---\emph{i.e.}\ with unit cells commensurate with the ($\sim$15\,\AA)$^3$ cell of the average structure. Each realisation relaxed to give a sensible approximant structure (\emph{e.g.}\ that shown in Figure 2E), with the ten possibilities giving equilibrium energies within a few kJ\,mol$_{\rm Zn}^{-1}$ of one another. The number of possible configurations grows very quickly with system size. For the $2\times2\times2$ case (32 formula units; 1696 atoms), there are too many possibilities to enumerate, so we focused on 20 representative realisations. These larger configurations spanned a wider range of energies and have more open structures [Figure 3A]. Since the enthalpic differences amongst these periodic TRUMOF-1 approximants are commensurate with solvent binding enthalpies in zinc carboxylate MOFs (\emph{e.g.}\ MOF-5),\cite{Hughes_2011,Brozek_2015} we expect that guest inclusion reduces the spread of energies in practice because loading would be higher in the less dense realisations. Hence we conclude that the configurational landscape accessible to TRUMOF-1---\emph{i.e.}\ representing different Truchet-tile arrangements---involves many thermally accessible minima, which helps to explain why synthesis at $\sim$400\,K generates a disordered, rather than ordered, configuration. Aperiodic networks will also be favoured on entropic grounds.

\begin{figure}
\begin{center} 
\includegraphics{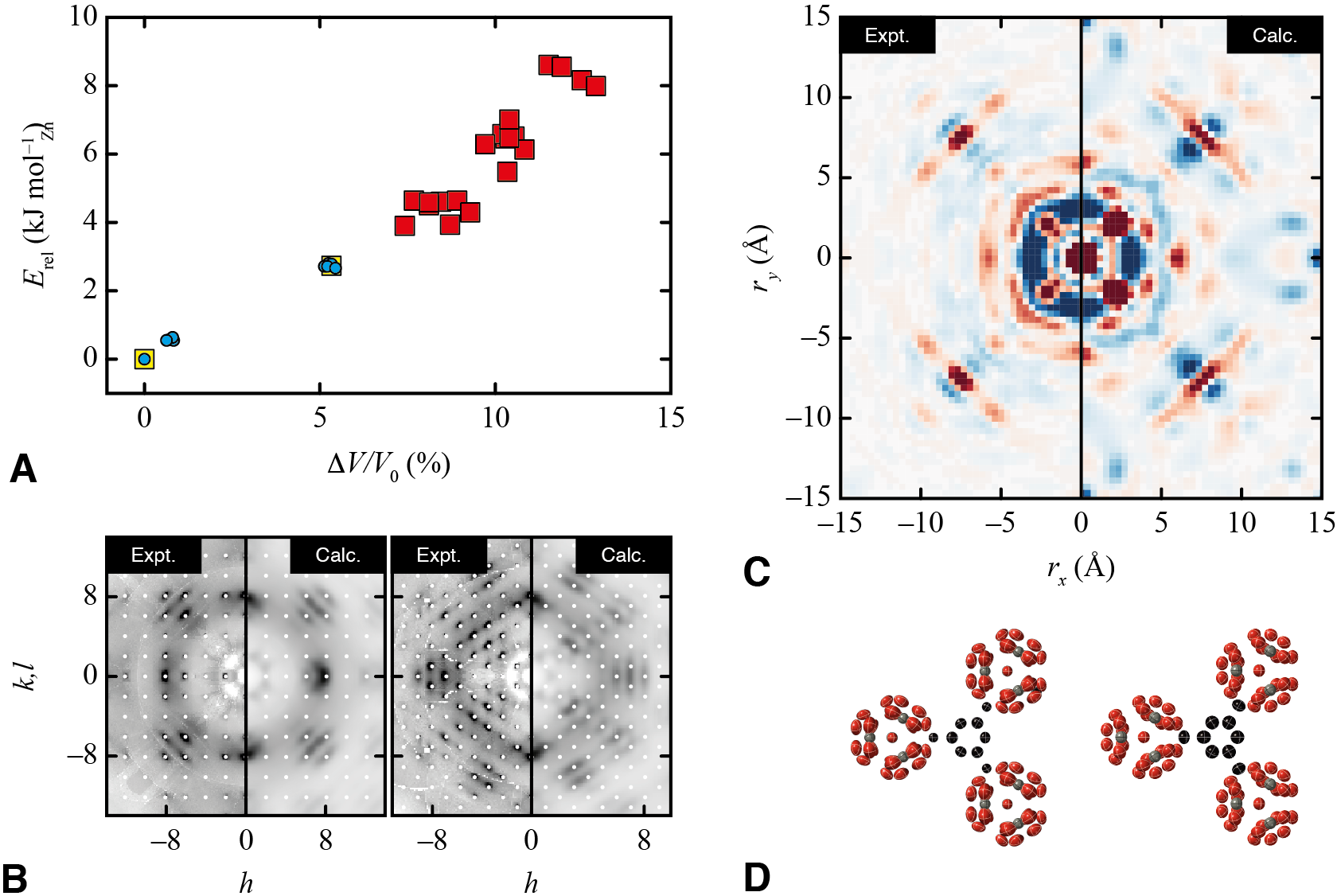}
\end{center}
\caption{\label{fig3}\footnotesize{\bf Computational and experimental validation of the TRUMOF-1 structure. A} The relative lattice energies of different $1\times1\times1$ (blue circles) and $2\times2\times2$ (red squares) approximants are within a few kJ\,mol$^{-1}_{\rm Zn}$ of one another. Also included are the energies for geometry-optimised $2\times2\times2$ supercells of two representative $1\times1\times1$ approximants (yellow squares). The more disordered configurations are those at higher energy, for which the molar volume is largest. {\bf B} Comparison of experimental and calculated single-crystal X-ray diffuse scattering in the (left) $(hk0)$ and (right) $(hhl)$ planes. The Bragg reflection positions are masked by white circles. {\bf C} Comparison between experimental and calculated $uv0$ cuts of the 3D-$\Delta$PDF. Positive correlations are shown in red and negative in blue. {\bf D} Comparison between the experimentally-derived crystallographic model for scattering density at the ligand site at 100\,K (left) and that obtained by projection of DFT-relaxed structures onto the same site (right). Ellipsoids are shown at 50\% probability.}
\end{figure}

The experimental measurement with greatest sensitivity to structural disorder and its correlation in crystalline materials is single-crystal diffuse scattering.\cite{Welberry_1985} A representation of the experimental X-ray diffuse scattering pattern of TRUMOF-1 is shown for the $(hk0)$ and $(hhl)$ scattering planes in Figure 3B: it is dominated by transverse-polarised diagonal features characteristic of correlated displacements.\cite{Welberry_2016} Because we do not find any strong temperature-dependence to the scattering, this contribution can be attributed predominantly to static displacements, rather than to thermal effects. The corresponding three-dimensional difference pair-distribution function (3D-$\Delta$PDF) captures the real-space correlations from which the diffuse scattering arises [Figure 3C].\cite{Weber_2012} The diffuse scattering and 3D-$\Delta$PDF calculated from our larger DFT-relaxed configurations are compared against experiment in Figure 3C,D; the close match indicates that the Truchet-tile model from which these configurations are generated captures the fundamental nature of correlated disorder in TRUMOF-1. The subtle differences we do observe can be attributed to the finite size of our DFT model---hence sharpness of features in reciprocal space---and the role of dynamical effects---most obviously affecting low-$|\mathbf r|$ features in the 3D-$\Delta$PDF. We anticipate even better agreement for larger approximants (were they computationally accessible), which would more closely approximate the aperiodic limit. As a further check against experiment, we show in Figure 3D that the distribution of atomic positions determined from configurational averaging over our DFT configurations is consistent with the anisotropic displacement parameters determined in our single-crystal X-ray diffraction refinements. Our model indicates that static disorder contributes approximately 0.1\,\AA$^2$ to atomic mean-squared displacements in the $F\bar43m$ average structure model; a similar value is obtained by extrapolating to 0\,K the experimental displacement parameters measured over 100--350\,K (see SI). Hence our Truchet-tile model rationalises at once both average and local structure measurements of the structure of TRUMOF-1.

The topology of TRUMOF-1 is unusual in a number of respects. The MOF network is a fully 6-connected net with nodes arranged on a periodic lattice (as in MOF-5) but its connectivity is aperiodic (in contrast to MOF-5). No two crystals have the same network connectivity---\emph{i.e.}\ they store different information---despite sharing the same statistical properties. In these senses, the system is intermediate to conventional crystalline MOFs, on the one hand, and amorphous MOF derivatives with continuous random network structures, on the other hand.\cite{Overy_2016,Bennett_2010} The topological aperiodicity of TRUMOF-1 may be important because it translates into a disordered pore network with specific characteristics.\cite{OrmrodMorley_2020} We know that guests are sensitive to this aperiodicity because loading TRUMOF-1 crystals with I$_2$ enhances the X-ray diffuse scattering contrast. A representative pore structure, extracted from one of our DFT configurations, is illustrated in Figure 4; the complex transport pathways one observes are qualitatively similar to those in gels\cite{Li_2020} and granular porous media\cite{Dullien_1992} and so may be useful in separations. We find good consistency between experimental adsorption behaviour and that calculated from Grand Canonical Monte Carlo simulations based on our structural model (see SI), and expect that application of the unusual pore-network structure of TRUMOF-1 will prove a fruitful area of future research. 

\begin{figure}
\begin{center} 
\includegraphics{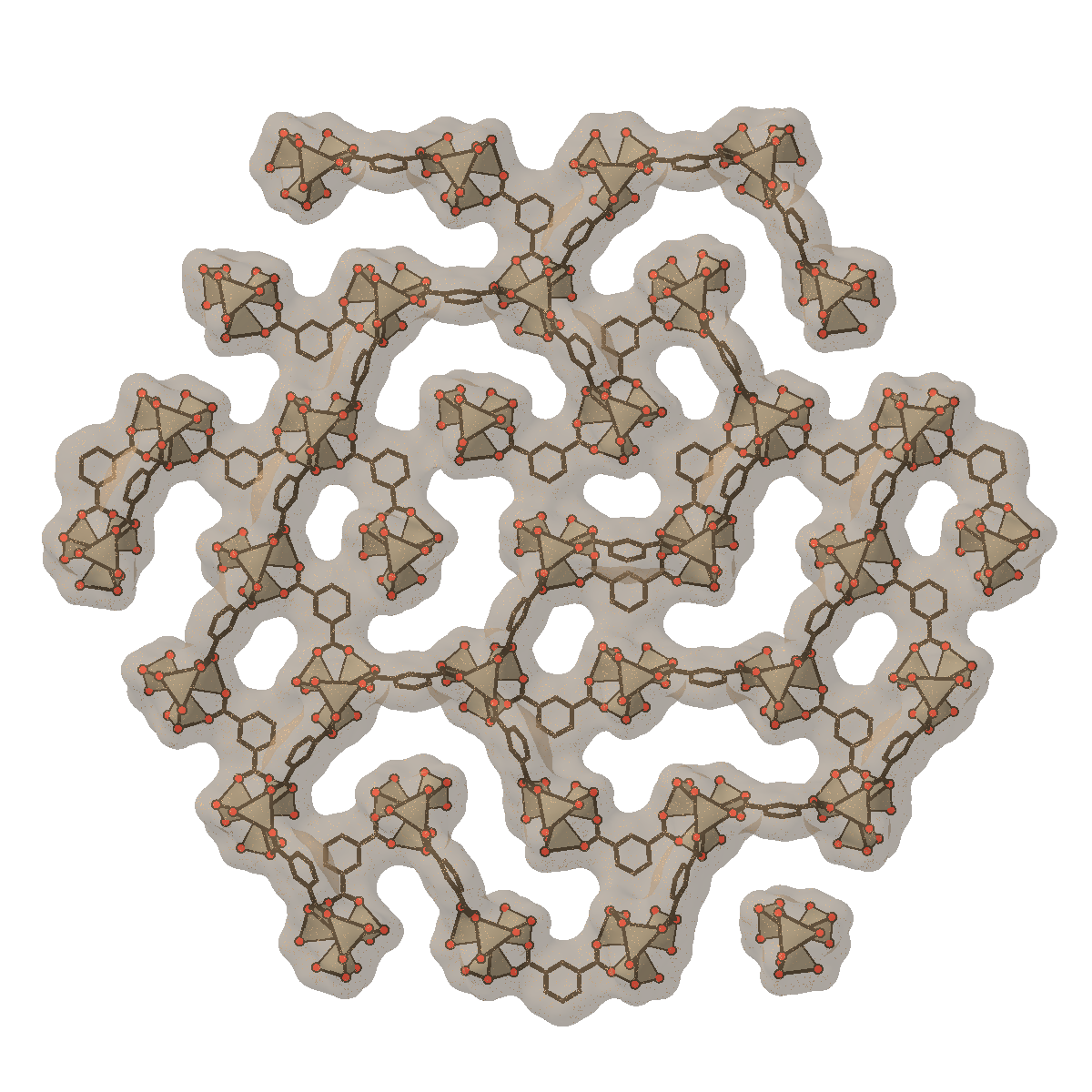}
\end{center}
\caption{\label{fig4}\footnotesize{\bf Disordered pore network structure.} A representative fragment of a representative TRUMOF-1 structure model, viewed down the $[111]$ crystal axis. Only those 1,3-bdc molecules that link clusters within this plane are shown. Each cluster is connected to six neighbouring clusters, some of which will be in planes above or below that shown here. The surface of the solvent-accessible pores is shown as a grey outline. Note that, despite the periodic arrangement of inorganic clusters, the pore network is disordered. There is a qualitative similarity to the network structure shown in Figure 1B.}
\end{figure}

There is no reason to expect that TRUMOF-1 should be the unique MOF with a Truchet-tile structure. Since every conventional periodic tiling can be modified to give Truchet tilings, suitable symmetry-lowering of inorganic and organic building units should allow targeted synthesis of alternative Truchet MOFs. Indeed it may prove to be the case that the structures of some previously-described MOFs\cite{Yang_2012b} and related systems\cite{Blunt_2008} with partially-occupied and/or complex structures might be re-interpretable within this alternative paradigm. It is probably no accident that TRUMOF-1 is assembled from an inorganic component predisposed to forming extended network structures\cite{Eddaoudi_2002} and an organic component usually employed in the formation of discrete molecular cages.\cite{Eddaoudi_2001} The concept of geometrically frustrating the competing structure-directing effects of two synthons to drive a complex state may generalise beyond this particular system.\cite{Zeng_2011} Our own preliminary investigations into synthesising additional TRUMOFs using differently substituted 1,3-bdc linkers always gave conventional crystalline products, but the synthetic conditions that favour Truchet-tile polymorphs may vary considerably from system to system. One way or the other, the discovery of even this one crystalline Truchet-tile material---MOF or otherwise---allows experimental investigation of the impact of aperiodic connectivity on collective materials properties. One might expect, for example, aperiodicity to frustrate elastic instabilities,\cite{Moshe_2018} or to dampen long-wavelength acoustic phonons---as occurs in superlattice heterostructures\cite{Venkatasubramanian_2001} and as required for optimising thermoelectric response.\cite{Gibson_2021} As such, Truchet-tile-inspired systems may provide a useful avenue for the design of a variety of functional systems.

\section*{References}

\bibliography{arxiv_2022_trumof}

\begin{addendum}
 \item The authors gratefully acknowledge financial support from the E.R.C. (Grant 788144). This work was supported by the Australian Research Council (to CJK; grants DP150104620, DP190103130) and also benefitted from access to the UK's ARCHER2 supercomputer through the Materials Chemistry Consortium, funded by EPSRC (EP/R029431). The authors acknowledge the facilities and the scientific and technical assistance of Sydney Analytical, a core research facility at The University of Sydney. 
\end{addendum}

\end{document}